\documentstyle[12pt]{article}
\begin{document}

\baselineskip 7 mm

\pagestyle{myheadings}

\begin{titlepage}

\begin{flushright}  McGill/93-28\\
hep-th/9401022
\end{flushright}

\begin{center}
\vskip 3em

{\LARGE Solution of matrix models} \\[1.5em]
{\LARGE in} \\[1.5em]
{\LARGE the DIII generator ensemble} \\

\vskip 3em
Harold Roussel\footnote{roussel@hep.physics.mcgill.ca}

\vskip 3em
Physics Department, McGill University\\
Ernest Rutherford Building\\
Montr\'eal, Qu\'ebec H3A 2T8\\
Canada\\

\begin{abstract}
In this work we solve two new matrix models, using standard
and new techniques.  The two models are based on matrix ensembles not
previously considered, namely the DIII generator ensemble.
It is shown that, in the double scaling limit, their
free energy has the same behavior as previous models
describing oriented and unoriented surfaces.  We also found an additional
solution for one of the models.
\end{abstract}

\end{center}

\end{titlepage}

\section*{Introduction}
Recent progress has been made in non-perturbative
string theory via matrix models, where one represents
triangulated random surfaces by large--\(N\) matrix
integrals.  By taking the so--called double scaling
limit, one gets some insight in the physics of the model,
associated with models of 2--D gravity.  In this paper we
look at two new such matrix models.

The reasons for studying
 them are that they were part of a classification scheme,
 but also they provided an opportunity to extend
 the techniques for solving matrix models, namely via skew-orthogonal
polynomials \cite{Maho:90}.  We also had to solve them
to compare with previous models to see if the physics
revealed by the free energy is the same.  Indeed, from its series
expansion, we can say if the model describes unoriented surfaces
in addition to oriented ones, and, by calculating an appropriate ratio,
we can say if it is the same as previous models.

Section one is devoted to a classification scheme
for matrix models within the context of symmetric spaces.
In section two and three, we show the calculations of two new
models in \( D=0 \).  Finally, we conclude with a discussion and a
comparison of the
two models' free energies, and a comparison with other models.

\section{Classification of matrix models}

We will now introduce a classification of matrix models
based on symmetric spaces.  This is an interesting class
of single--matrix models for which the reduction of the
matrix integral allows for a solution by a polynomial method.

Symmetric spaces are cosets of the form \(G/K\).  Technically
speaking, those that we are considering are the simply
connected Riemannian globally symmetric spaces \cite{Pere:83,Helg,Gilm}.
The classical types are listed in table~\ref{tablech1}.
\begin{table}[p]
\begin{center}
\begin{tabular}{|l|l|l|l|l|} \hline
Cartan's & & & System of & \\
notation & \(X=G/K\) & Rank & restricted & Multiplicity \\
& & & roots & \\ \hline
A I & $SL(n,R)/SO(n)$ & $n-1$ & $A_{n-1}$ & $m_{\alpha} = 1$ \\
A II & $SU(2n)/Sp(n)$ & $n-1$ & $A_{n-1}$ & $m_{\alpha} = 4$ \\
A III & $SU(p,q)/S(U(p) \times U(q))$ & $n$=min($p$,$q$) & $p=1$,
 $C_{n}$ & $m_{\alpha} = 2$,
$m_{2 \beta} = 1$ \\
& & & $p>q$,$B C_{n} $ & $m_{\alpha} = 2$, $m_{2 \beta} = 1$ \\
& & & & $m_{\beta} = 2(p-q) $ \\
BD I & $SO(p,q)/SO(p) \times SO(q)$ & $n$=min($p$,$q$) & $p=q$,$D_{n}$ &
 $m_{\alpha} = 1$ \\
& & & $p>q$,$B_{n} $ & $m_{\beta} = (p-q)$ \\
& & & & $m_{\alpha} = 1$ \\
D III & $SO(2n)/U(n)$ & $n/2$ & $n=2 k$,$C_{k}$ & $m_{\alpha} = 4$,
$m_{2 \beta} = 1$ \\
& & & $n=2 k +1$ & $m_{\alpha} = 4$, $m_{2 \beta} = 1$ \\
& & & $B C_{k}$ & $m_{\beta} = 4 $ \\
C I & $Sp(n,R)/U(n)$ & $n$ & $C_{n}$ & $m_{\alpha} = 1$ \\
C II & $Sp(p,q)/Sp(p) \times Sp(q)$ & $n$=min($p$,$q$) & $p=q$,
 $C_{n}$ & $m_{\alpha} = 4$, $m_{2 \beta} = 3$ \\
& & & $p>q$, $B C_{n}$ & $m_{\alpha} = 4$, $m_{2 \beta} = 3$ \\
& & & & $m_{\beta} = 4(p-q)$ \\ \hline
\end{tabular}
\end{center}
\caption{Classification of the symmetric spaces.}
\label{tablech1}
\end{table}
The two matrix models, in sections two and three, are based on the DIII
generator ensemble.
They correspond, for section two, to \(n\) odd, and for section three,
to \(n\) even.

There are a few ways to reduce a matrix integral,
depending on the type of matrix or the matrix ensemble.  For example, for the
Hermitian matrix model, one can simply diagonalize the matrix as it is done
in Metha's book \cite{Meth}.  The result is that we get an integral over
the eigenvalues with a Jacobian in terms of these eigenvalues.  The idea
behind this reduction is to simplify the matrix integral.

The symmetric spaces introduced in the previous subsection give us a
systematic approach to identify the matrix ensemble for which a similar
reduction is possible.  For these spaces, however, the
procedure of reducing the matrix integral is more complex.
  An analogy can be used to understand
how the Jacobian and the set of integration parameters arise.
  Consider the volume element in flat space written
in cartesian coordinates and in spherical coordinates, \(\prod_{i=1}^{N}
dx_{i} = d \omega\,dr\,r^{N-1} \).  Spherical symmetry would allow
us to integrate out the angular part \( d \omega \) with \( r^{N-1} \) being
the Jacobian and \( r\) the integration variable.
  For a symmetric space (matrix ensemble) the Haar measure
can be decomposed in the same way, with the symmetry provided by the
associated group.  In doing so we get the Jacobian and also a set of
parameters (which are not necessarily the eigenvalues of the matrices).
The example above can be regarded as the measure for the
generator ensemble of the coset \(SO(N+1)/SO(N)\).  The parameter of
interest is \(r\) and \(d \omega \) is the measure on
\( SO(N) \), which yields a ``trivial" integral if the original
integrand is spherically symmetric.  Such vector models were studied in
\cite{Rav:91}.
  For the technical details the reader is referred to Helgason's
book~\cite{Helg}.  Let us just say here that there is a relation
between the multiplicities of the roots and the exponents in the Jacobian.
Table~\ref{tablech12} gives some symmetric spaces and their
 associated Jacobian.
Many of these models have been previously
 considered, including A I \cite{Neub:91,Mart:90},
corresponding to real symmetric matrices, A II \cite{Mart:90,Myerp:90},
corresponding to quaternionic real self dual
 matrices, as well as A III , BD I, and C II \cite{Rav:91,Myer:92}.

It should be noted here that matrix models with Hermitian,
antisymmetric, and quaternionic real self dual matrices, are
 not part of this classification.  In fact,
they are generator ensembles for the
 classical groups \( U(n)\),\(SO(n)\), \(Sp(2n)\),
respectively.

In the present paper we consider the D III generator ensemble,
so only C I remains to be studied.

\begin{table}[p]
\begin{center}
\begin{tabular}{|c|c|c|} \hline
Cartan's & System of & Jacobian for \\
notation & restricted roots & generator ensemble \\ \hline
A I & \( A_{n-1} \) & \( \displaystyle{ \prod_{i<j} |x_{i} - x_{j} | } \) \\
A II & \( A_{n-1} \) & \( \displaystyle{\prod_{i<j} |x_{i} - x_{j} |^{4}} \) \\
A III & \( p=1 \) , \( C_{n} \) & \( \displaystyle{\prod_{i<j}
(x_{i}^{2} - x_{j}^{2} )^{2} \prod | x_{i}|^{2(1-q)+1}} \) \\
& \( p>q \) , \( B C_{n} \) & \( \displaystyle{\prod_{i<j} (x_{i}^{2} -
x_{j}^{2})^{2} \prod | x_{i}|^{2(p-q)+1}} \) \\
BD I & \( p=q \) , \( D_{n} \) & \( \displaystyle{\prod_{i<j}
|x_{i}^{2} - x_{j}^{2} | } \) \\
& \( p>q \) , \( B_{n} \) & \( \displaystyle{\prod_{i<j} |x_{i}^{2} -
x_{j}^{2}| \prod | x_{i}|^{p-q}} \) \\
C I & \( C_{n} \) & \( \displaystyle{ \prod_{i<j} |x_{i}^{2} -
 x_{j}^{2}| \prod | x_{i}} | \) \\
C II & \( p=q \) , \( C_{n} \) & \( \displaystyle{\prod_{i<j}
(x_{i}^{2} - x_{j}^{2} )^{4} \prod | x_{i}|^{3}} \) \\
& \( p>q \) , \( B C_{n} \) & \( \displaystyle{\prod_{i<j} (x_{i}^{2} -
x_{j}^{2})^{4} \prod | x_{i}|^{4(p-q)+3}} \) \\
D III & \( n=2 k \) , \(C_{k} \) & \( \displaystyle{\prod_{i<j}
(x_{i}^{2} - x_{j}^{2})^{4} \prod | x_{i}} | \) \\
& \( n=2 k +1 \), \( B C_{k} \) & \( \displaystyle{\prod_{i<j}
(x_{i}^{2} - x_{j}^{2})^{4} \prod | x_{i}^{5}} | \) \\ \hline
\end{tabular}
\end{center}
\caption{Symmetric spaces and their Jacobians.}
\label{tablech12}
\end{table}

\section{First model -- n odd}
We will first study the double--scaling limit of the D III generator
ensemble with \(n\) odd (see table \ref{tablech12}).  The matrix
integral always takes the form,
\begin{equation}
{\cal Z} = \int dM e^{- \beta\,Tr\,V(M)}.
\label{eq:matrixintegral}
\end{equation}
The Jacobian for this model is
\begin{equation}
{\cal J}=\prod _{i} ^{N} |x_{i} ^{5}| \prod _{i<j} ^{N} (x_{i} ^{2} -
x_{j} ^{2})^{4},
\end{equation}
and we are going to make the analysis with the potential \( V=a x^{2}
+ b/2\,x^{4} \).  The partition function
can then be written as follows,
\begin{equation}
{\cal Z}_{N} = 2^{-N}\int _{0} ^{\infty} \prod _{i} ^{N} dy_{i} \prod
_{i} ^{N} y_{i}^{2}
\prod _{i<j} ^{N} (y_{i} - y_{j})^{4} e^{- \beta (a y_{i} +b/2 y_{i} ^{2}
)}, \label{eq:eq334}
\end{equation}
where we made the change of variables \(x^{2} =y \).
For convenience of evaluation we will
rewrite it as a determinant \cite{Meth}.  We have,
\begin{equation}
\prod _{i} ^{N} y_{i}^{2} \prod _{i<j} ^{N} (y_{i} - y_{j})^{4} =
\det \left[
\begin{array}{cccccc}
P_{0}(y_{1}) & P_{1}(y_{1}) & & \ldots & & P_{2N-1}(y_{1}) \\
y^{2} \partial_{y} P_{0}(y_{1}) & y^{2} \partial_{y} P_{1}(y_{1}) & & \ldots &
&
y^{2} \partial_{y} P_{2N-1}(y_{1}) \\
P_{0}(y_{2}) & P_{1}(y_{2}) & & \ldots & & P_{2N-1}(y_{2}) \\
 & & & & &  \\
\vdots & \vdots & & & & \vdots \\
 & & & & &  \\
y^{2} \partial_{y} P_{0}(y_{N}) & y^{2} \partial_{y} P_{1}(y_{N}) & & \ldots &
&
y^{2} \partial_{y} P_{2N-1}(y_{N})
\end{array}
\right].
\label{eq:determ}
\end{equation}
The \(P\)'s are the usual orthogonal polynomials on the half-line, defined as
\begin{eqnarray}
P_{k}(y)=y^{k} + l.o. & P_{0}(y)=1 \\
\int_{0}^{\infty} d \mu\,P_{k}(y) P_{l}(y) = h_{k} \delta _{l,k} & d \mu=
dy\,e^{- \beta V(y)}
\end{eqnarray}
and from which we derive the following recursion relation,
\begin{equation}
y P_{k}= P_{k+1} + S_{k} P_{k} + R_{k} P_{k-1}, \label{eq:eq338}
\end{equation}
with \(R_{k} = h_{k} / h_{k-1} \).

In order to solve for the free energy, we will need to know, first,
the solutions for \(R\) and \(S\).
By considering recursion relations for \(\partial_{y} P_{k}\) and
\( y \partial_{y} P_{k}\), one
finds a set of two self-consistent equations in terms of \(R\), \(S\),
 \(k\), \(a\), \(b\), and \( \beta \) \cite{Morr:91},
\begin{equation}
{4 k+1 \over \beta} = a S_{2k} + b (R_{2k} + R_{2k+1} + S_{2k}^{2})
\label{eq:equation2}
\end{equation}
and
\begin{equation}
(a+b S_{2k})(a+b S_{2k-1}) R_{2k} = {1 \over 4} [S_{2k} (a+b S_{2k}) +
b(R_{2k+1}-R_{2k}) - {1 \over \beta}]^{2}. \label{eq:equation3}
\end{equation}

We used \(2k\) indices because in the final recursion
relation, we will need to know the solutions of \(R_{2N+l}\)
and \(S_{2N+l}\).  In the planar approximation, and choosing \(
\lambda_{c} = (k/ \beta)_{c} = 1/4 \)
and \( R_{c}=1 \) (together with the criticality condition \(\partial
_{S} \lambda =0\) ), we find all the critical values,
\begin{equation}
\begin{array}{ccc}
\lambda_{c}= 1/4 & R_{c} = 1 & S_{c}= 2 \\
& a=1 & b=-1/6.
\end{array}
\end{equation}

We choose the scaling solutions to be,
\begin{eqnarray}
S_{2N+l} & = & 2 (1- \beta ^{- \mu} \exp(-{l \over 2} \beta ^{- \nu} {\partial
\over
\partial t}) f ) \\
R_{2N+l} & = & 1 - \beta ^{-\mu} \exp(-{l \over 2} \beta ^{-\nu} {\partial
\over
\partial t}) ( g_{0} + \beta ^{-\nu} g_{1} + g_{2} \beta^{-2 \nu} + \ldots )
\end{eqnarray}
where \(t\) is defined by \( t=(\beta/4 - N) \beta ^{- \nu} \).  We write
\(N\) instead of \(k\) because we are only interested in the large N behavior
of \(R\) and \(S\) (and so we replace \(k\) by \(N\)
 in our recursion relations as well).  In the \(R\)
ansatz we added some more degrees of freedom by expanding the \(g\)
function in powers of \(\beta^{-\nu}\).  This will be necessary to get
consistent solutions beyond leading order in \( \beta^{-\nu} \).
  Doing the same thing for \(f\) would only yield
redundant equations.

After inserting in the recursion relations, and working out the
lowest order equations, we see that a consistent
solution requires \( \mu = 2 \nu \).  For the two recursion relations,
the results are given, order by order, in table
\ref{tablech2}.

\begin{table}[p]
\begin{center}
\begin{tabular}{|c|c|c|} \hline
order & eq.(\ref{eq:equation3}) & eq.(\ref{eq:equation2}) \\ \hline
& & \\
\( \beta^{0} \) & \( 4/9=4/9 \) & 1=1 \\
& & \\
\( \beta^{-\mu} \) & \( g_{0}=2 f \) & \( g_{0}=2 f \) \\
& & \\
\( \beta^{-\mu -\nu} \) & \( g_{1} = f'/2 \) & \( g_{0}'=4 g_{1} \) \\
& & \\
\( \beta^{-2 \mu} = \beta^{-\mu -2 \nu} \) & \( g_{2} = 1/16 f'' - f^{2} \)
& \(4t=2/3\,f^{2}-1/3\,g_{2}+1/12\,g_{1}'-1/48\,g_{0}''\) \\
& & \\ \hline
\end{tabular}
\end{center}
\caption{Lowest order solutions of the two recursion relations for
\(R_{2N+l}\) and \(S_{2N+l}\).}
\label{tablech2}
\end{table}

{}From eq.(\ref{eq:equation2}), we obtain the values of the exponents,
\( \nu=1/5 \), and \( \mu=2/5 \), which is consistent with our
previous relation between \( \mu \) and
\( \nu \).  Also,
using the relations between the \(g\) coefficients and \(f\),
we get, at order \( \beta^{-\nu -2 \nu} \)
 in (\ref{eq:equation3}), a differential equation for \(f\),
\begin{equation}
4t= f^{2} -1/48\,f'' \label{eq:painleve}
\end{equation}
which is the well-known Painlev\'e I equation, upon renormalization of
\(t\).
At order \( \beta^{-\mu} \) and \( \beta^{-\mu-\nu} \), we simply get
equalities (e.g. \(0=0\)), which come from our criticality requirement (
first derivative of \(\lambda\) with respect to \(S\) is zero).

To solve the model, one must still solve for the double--scaling
limit of the entire partition function.  For this, it is clear that we
have to find a recursion relation for \( y^{2} \partial _{y} P_{n}(y) \)
(as appears in the determinant \ref{eq:determ}).
After some algebraic manipulations, one finds
\begin{eqnarray}
y^{2} \partial_{y} P_{n} = & n P_{n+1} + \left(
\displaystyle{\beta \over 2} C_{n,n} -
S_{n} \right) P_{n} + (\beta C_{n-1,n} -n -1) R_{n} P_{n-1} + \nonumber \\
 & \beta C_{n-2,n} R_{n} R_{n-1} P_{n-2} + \beta C_{n-3,n} R_{n}
R_{n-1} R_{n-2} P_{n-3},
\end{eqnarray}
where, for simplicity we wrote \( C_{k,l} = \int d \mu P_{k} P_{l} y^{2} V'(y)
\).
With this recursion relation, the partition function is found to be,
\begin{eqnarray}
{\cal Z}_{N+1} & = (N+1) (<P'_{2N}, P_{2N+1}> - <P'_{2N+1}, P_{2N}>)
{\cal Z}_{N} +  N(N+1) \nonumber \\
& <P'_{2N-2}, P_{2N+1}>(<P'_{2N-1}, P_{2N}> - <P'_{2N}, P_{2N-1}>)
{\cal Z} _{N-1}  \nonumber \\
& - N(N+1)(N-1) <P'_{2N-2}, P_{2N+1}> <P'_{2N-3}, P_{2N}> \nonumber \\
& <P'_{2N-4}, P_{2N-1}> {\cal Z} _{N-2}
-(N+1) <P'_{2N-1}, P_{2N+1}> Y_{N},
\label{eq:zequation1}
\end{eqnarray}
where \(P' = y^{2} \partial_{y} P \). \(Y_{N}\) is an auxilliary partition
function with a determinant similar to \(Z_{N}\) but where the last
and third--to--last columns were removed.  It was introduced to avoid
an infinite number of terms in the previous relation, and satisfies
\begin{equation}
Y_{N} = N <P'_{2N-2}, P_{2N}> {\cal Z}_{N-1} - N <P'_{2N-3}, P_{2N}>
Y_{N-1}.
\label{eq:yequation1}
\end{equation}
We now define, for later convenience (we want to have
a smooth planar limit as \(N \rightarrow \infty\)~), the following two ansatz,
\begin{equation}
\begin{array}{lcr}
{\displaystyle W_{N} = { {\cal Z}_{N} \over {\cal Z}_{N-1} N \beta b
 h_{2N-1}}} & & {\displaystyle X_{N}={ Y_{N} \over {\cal Z}_{N-1}
 N \beta b h_{2N-1}} },
\end{array} \label{eq:anstz}
\end{equation}
and rewrite eqs.(\ref{eq:zequation1}) and (\ref{eq:yequation1}) in terms of
polynomials in \(W\) and \(X\).  After some algebraic manipulations,
and expanding the brackets,
we get
\begin{eqnarray}
& {\displaystyle W_{N+1} W_{N} W_{N-1} - {W_{N} W_{N-1} \over b}
 ( a (S_{2N+1}+ S_{2N})
+ b (R_{2N}+R_{2N+1} +R_{2N+2} +S_{2N+1}^{2} } \nonumber \\
& {\displaystyle +S_{2N}^{2}+S_{2N+1} S_{2N})
- {4N \over \beta} - {2 \over \beta} ) + {X_{N} W_{N-1} \over b} \left(
a+b (S_{2N+1}+S_{2N}+S_{2N-1}) \right) } \nonumber \\
& {\displaystyle - W_{N-1} R_{2N} (R_{2N+1}
 +R_{2N} +R_{2N-1} +S_{2N-1}^{2}
+S_{2N}^{2} +S_{2N-1} S_{2N} )  } \nonumber \\
& {\displaystyle - {W_{N-1} R_{2N} \over b} \left( a
 (S_{2N} +S_{2N-1}) - {4N \over \beta}
\right) + R_{2N} R_{2N-1} R_{2N-2} } =0 \label{eq:wequation2} \\
& \nonumber \\
& {\displaystyle X_{N+1} W_{N} - {R_{2N+2} W_{N} \over b}
 (a+b (S_{2N+2} +S_{2N+1}+
S_{2N})) +R_{2N+2} X_{N} =0 }. \label{eq:xequation2}
\end{eqnarray}
In the planar limit, and using the previously found critical values
for \(R\), \(S\), etc, we find that
the critical values are \(W_{c}=-1\) (triple root of
\ref{eq:wequation2}), while \(X_{c}\) remains undetermined.
\(X\) it will be determined below when solving the recursion
relations.

To solve them, we choose the following ansatz,
\begin{eqnarray}
W_{N+l} & = & -1 + \beta ^{-\rho} \exp(-l \beta ^{-\nu} {\partial \over
\partial t}) (h_{0} + \beta ^{-\nu} h_{1} + \beta ^{-2 \nu} h_{2}+ \ldots) \\
X_{N+l} & = & X_{c} - \beta ^{-\sigma} \exp(-l \beta ^{-\nu} {\partial
\over \partial t}) (k_{0}+ \beta^{-\nu} k_{1}+ \beta^{-2 \nu} k_{2}+\ldots).
\end{eqnarray}
After insertion in eqs.~(\ref{eq:wequation2}) and (\ref{eq:xequation2})
 (at \( \beta ^{-2 \nu} \) order ), and using \( \phi = \mu = 2/5 \),
\( \sigma = \rho = \nu =
1/5 \), we find the results given in table~\ref{tablech3}.
\begin{table}[p]
\begin{center}
\begin{tabular}{|c|c|c|} \hline
order & eq.(\ref{eq:wequation2}) & eq.(\ref{eq:xequation2}) \\ \hline
& & \\
\( \beta^{-1/5} \) &
 \( 0 \) & \( h_{0} X_{c} = 0 \) \\
& & \\
\( \beta^{-2/5} \) &
 \( 6 f X_{c}=0 \) &
\( -6 f - g_{0} X_{c} + h_{1} X_{c} - h_{0} k_{0} - k_{0}'= 0 \) \\
& & \\
\( \beta^{-3/5} \) &
 \( -3 g_{0}' -6 g_{0} h_{0} +h_{0}^{3} +3 h_{0} h_{0}' \) &
\( 3 f' +6 f h_{0} +g_{0} k_{0} -h_{1} k_{0}+ h_{0} k_{0}'-h_{0} k_{1} \) \\
& \(+h_{0}''-6 f k_{0} -6 f h_{0} X_{c}=0 \) & \( k_{0}''/2 -g_{1} X_{c}
+g_{0}' X_{c} +h_{2} X_{c} -k_{1}'=0 \) \\
& & \\ \hline
\end{tabular}
\end{center}
\caption{Lowest order solutions of the two recursion relations for the DIII
-- odd n -- matrix model.}
\label{tablech3}
\end{table}
We see that a consistent solution requires that \(X_{c}=0\).  In doing so,
we get two differential equations defining \(h_{0}\) and \(k_{0}\),
\begin{eqnarray}
0 & = & 6 f +h_{0} k_{0} +k_{0}' \\
0 & = & -6 f' -12 f h_{0} +h_{0}^{3} +3 h_{0} h_{0}' +h_{0}'' -6 f k_{0}.
\end{eqnarray}
We used the fact that \(g_{0}=2 f\) (table \ref{tablech2}).  Using the
known solution for \(f\),
\begin{equation}
f= 2 t^{1/2} - {1 \over 384} t^{-2} - {49 \over 589824} t^{-9/2} - \ldots,
\end{equation}
and power series solution
for \(h_{0}\) and \(k_{0}\),we find a set of algebraic equations that we can
solve.  We finally end up with four solutions,
\begin{eqnarray}
h_{0} & = & \pm 2 \sqrt{3} t^{1/4} + {7 \over 8} t^{-1} \mp {5 \sqrt{3} \over
384} t^{-9/4} + \ldots \label{eq:sols1} \\
k_{0} & = & \mp 2 \sqrt{3} t^{1/4} + {9 \over 8} t^{-1} \pm {5 \sqrt{3}
\over 384} t^{-9/4} + \ldots \nonumber
\end{eqnarray}
or
\begin{eqnarray}
h_{0} & = & \pm 2 \sqrt{3} t^{1/4} - {1 \over 8} t^{-1} \mp {5 \sqrt{3} \over
384} t^{-9/4} + \ldots \label{eq:sols2} \\
k_{0} & = & \mp 2 \sqrt{3} t^{1/4} + {1 \over 8} t^{-1} \pm {5 \sqrt{3}
\over 384} t^{-9/4} + \ldots. \nonumber
\end{eqnarray}
It turns out that we will only need \(h_{0}\) in the solution of the free
energy.  Using the following relation for large \(N\),
\begin{equation}
\exp{(- \partial_{N}^{2} F)} = {{\cal Z}_{N+1} {\cal Z}_{N-1} \over
{\cal Z}_{N}^{2} } = R_{2 N+1} R_{2 N} {W_{N+1} \over W_{N} },
\end{equation}
the second derivative of the free energy is found to be
\begin{equation}
F'' \simeq 4 f-h_{0}'.
\end{equation}
Using the solution for \(f\) and the solution (\ref{eq:sols2}) for
\(h_{0}\),
we find,
\begin{equation}
F'' = 8 t^{1/2} \mp {\sqrt{3} \over 2} t^{-3/4} - {13 \over 96} t^{-2} +
\ldots.
\end{equation}
In order to compare this result with other matrix models, we need a
quantity independent of the scale of \(t\) and \(F''\).  Such a quantity is
the product of the first and the third coefficient in the series for \(F''\)
divided by the square of the second coefficient.  Here we find \(-13/9\),
which is the same result
 as in \cite{Neub:91,Mart:90} although the matrix ensemble in these
papers is different.  From previous solutions this is the expected ratio for
the free energy of pure 2D quantum gravity with oriented and unoriented
surfaces.
Doing the same with (\ref{eq:sols1}), we get,
\begin{equation}
F'' = 8 t^{1/2} \mp {\sqrt{3} \over 2} t^{-3/4} + {83 \over 96} t^{-2} +
\ldots.
\end{equation}
In that case, the universal ratio \(c_{0} c_{2}/ c_{1}^{2} \) yields
\(83/9\), which differs from any known matrix models.  One surprising
result is that our two solutions differ
only by \(t^{-2}\) (verified up to 15th order of \(F''\) ),
which is the term corresponding to the torus and the Klein bottle.

\section{Second model -- n even}

We start again from the matrix integral (\ref{eq:matrixintegral}).  But now,
the
Jacobian takes the form,
\begin{equation}
{\cal J}=\prod _{i} ^{N} |x_{i}| \prod _{i<j} ^{N} (x_{i} ^{2} -
x_{j} ^{2})^{4}.
\end{equation}
We are doing the analysis, as usual, with the potential \( V=a x^{2}
+ b / 2\,x^{4} \).
 The partition function can then be written as follows,
\begin{equation}
{\cal Z}_{N} = 2^{-N} \int _{0} ^{\infty} \prod _{i} ^{N} dy_{i}
\prod _{i<j} ^{N} (y_{i} - y_{j})^{4} e^{- \beta (a y_{i} +b/2 y_{i} ^{2}
)}.
\end{equation}
We used, again, the substitution \(x^{2}=y \).  The complicated part
is the Jacobian.  For convenience of evaluation we
rewrite it as a determinant \cite{Maho:90}.  We have,
\begin{equation}
\prod _{i<j} ^{N} (y_{i} - y_{j})^{4} =
\det \left[
\begin{array}{cccccc}
Q_{0}(y_{1}) & Q_{1}(y_{1}) & & \ldots & & Q_{2N-1}(y_{1}) \\
\partial_{y} Q_{0}(y_{1}) & \partial_{y} Q_{1}(y_{1}) & & \ldots & &
\partial_{y} Q_{2N-1}(y_{1}) \\
Q_{0}(y_{2}) & Q_{1}(y_{2}) & & \ldots & & Q_{2N-1}(y_{2}) \\
 & & & & &  \\
\vdots & \vdots & & & & \vdots \\
 & & & & &  \\
\partial_{y} Q_{0}(y_{N}) & \partial_{y} Q_{1}(y_{N}) & & \ldots & &
\partial_{y} Q_{2N-1}(y_{N})
\end{array}
\right],
\end{equation}
where the \(Q\)'s are Metha's skew-orthogonal polynomials,
\begin{equation}
Q_{i}(y)=y^{i} + l.o.,
\end{equation}
and
\begin{equation}
<Q_{i},Q_{j}>_{Q} \equiv {1 \over 2} \int dy\,e^{-\beta V(y)}
(Q_{i} Q_{j}' - Q_{i}' Q_{j}) = q_{[i/2]} z_{i j},
\end{equation}
with
\begin{eqnarray}
z_{2i,2i+1} & = & 1 \\
z_{2i+1,2i} & = & -1 ,
\end{eqnarray}
all others being 0.  With these definitions we can easily evaluate \(
{\cal Z}_{N} \),
\begin{equation}
{\cal Z}_{N}=2^{-N} N! \prod _{i=0} ^{N-1} q_{i}. \label{eq:partfun}
\end{equation}

Here we used skew-orthogonal polynomials because with the orthogonal ones,
\( {\cal Z}_{N} \) cannot be evaluated (i.e. gives an
 infinite recursion relation
for \( \partial P \) ).  On the other
hand, with \(Q\) polynomials, \( {\cal Z}_{N} \) is easy to find,
 but the problem is
to establish recursion relations for them and the \(q\)'s.
  Indeed, the only know recursion
relation is infinite \cite{Maho:90}.  The approach that we will follow
is to relate the \(Q\) polynomials with the \(P\) polynomials
 (these are the usual
orthogonal polynomials for which recursion relations are well-known, or
at least, easy to find).

We start with the general expansion of the \(P\)'s in term of the \(Q\)'s,
\begin{eqnarray}
P_{2i} & = & Q_{2i} + \omega_{i 1} Q_{2i-1} + \omega_{i 2} Q_{2i-2} + \ldots\\
P_{2i+1} & = & Q_{2i+1} + \xi_{i 1} Q_{2i} + \xi_{i 2} Q_{2i-1} + \ldots.
\end{eqnarray}
In order to find odd and even \( \xi \)'s and \( \omega \) 's, we
consider four \(Q\) products.  In the calculation we take the \(Q\)
product of a \(P\) polynomial and a \(Q\) polynomial.
After using the expansion of \(P\)'s in terms of \(Q\)'s,
and rewriting the \(Q\) product in a \(P\) product (orthogonal
polynomial relation), we get the following relations,
\begin{eqnarray}
\xi_{i,2j} q_{i-j} & = & {1 \over 2} <\beta V' Q_{2i-2j}, P_{2i+1}>_{P}
- {1 \over 2} Q_{2i-2j}(0) P_{2i+1}(0) \label{eq:qproda} \\
\omega_{i,2j-1} q_{i-j} & = & {1 \over 2} <\beta V' Q_{2i-2j}, P_{2i}>_{P}
 - {1 \over 2} Q_{2i-2j}(0) P_{2i}(0) \label{eq:qprodb} \\
-\omega_{i,2j} q_{i-j} & = & {1 \over 2} <\beta V' Q_{2i-2j+1}, P_{2i}>_{P}
 - {1 \over 2} Q_{2i-2j+1}(0) P_{2i}(0) \nonumber \\
& & - {1 \over 2} (2i+1) \delta_{j 0} h_{2i} \label{eq:qprodc} \\
-\xi_{i,2j+1} q_{i-j} & = & {1 \over 2} <\beta V' Q_{2i-2j+1}, P_{2i+1}>_{P}
 - {1 \over 2} Q_{2i-2j+1}(0) P_{2i+1}(0). \label{eq:qprodd}
\end{eqnarray}
Considering the above equations for specific values of \(j\), and defining
the following quantities which have a smooth planar limit,
\begin{equation}
\begin{array}{ccc}
\displaystyle{W_{i}={q_{i} \over \beta h_{2i}}} &
\displaystyle{ X_{i}={Q_{2i}(0) \over P_{2i}(0)}} &
\displaystyle{ Y_{i}={Q_{2i+1}(0) \over P_{2i}(0)}} \\
& \displaystyle{Z_{k}={P_{k+1}(0) \over P_{k}(0)}} &
\displaystyle{ A_{k}={P_{k}(0)^{2} \over \beta
h_{k}}},
\end{array} \label{eq:scal}
\end{equation}
we finally get,
\begin{eqnarray}
{W_{i} \over R_{2i+1}} & = & {b \over 2} - {1 \over 2} {X_{i} A_{2i+1}
\over Z_{2i}} \label{eq:equa24} \\
{\xi_{i 2} W_{i-1} \over R_{2i+1} R_{2i} R_{2i-1}} & = &
- {1 \over 2} {X_{i-1} A_{2i+1} \over Z_{2i} Z_{2i-1} Z_{2i-2}}
 \label{eq:equa25} \\
0 & = & a+b S_{2i} - \omega_{i 1} b - X_{i} A_{2i} \label{eq:equa26} \\
{\omega_{i 1} W_{i-1} \over R_{2i} R_{2i-1}} & = &
- {1 \over 2} {X_{i-1} A_{2i} \over Z_{2i-1} Z_{2i-2}} \label{eq:equa27} \\
-2 W_{i} & = & b R_{2i+1} - \xi_{i 1}(a+b S_{2i}) + (\xi_{i 1}
\omega_{i 1}- \xi_{i 2}) b - Y_{i} A_{2i} - {(2i+1) \over \beta}
 \label{eq:equa28} \\
{- \omega_{i 2} W_{i-1} \over R_{2i} R_{2i-1}} & = &
{b \over 2} - {1 \over 2} {Y_{i-1} A_{2i} \over Z_{2i-1} Z_{2i-2}}
 \label{eq:equa29} \\
{-2 \xi_{i 1} W_{i} \over R_{2i+1}} & = & a+b S_{2i+1} - b \xi_{i 1}
- {Y_{i} A_{2i+1} \over Z_{2i}}. \label{eq:equa30}
\end{eqnarray}
In the planar limit, and using critical values found in the previous
section, we find,
\begin{equation}
\begin{array}{ccccc}
b \rightarrow -1/6 & \lambda=i/\beta \rightarrow 1/4 &
S \rightarrow 2 & Z \rightarrow -1 & \omega_{i,n} \rightarrow \omega_{n} \\
& a \rightarrow 1 & R \rightarrow 1 & A \rightarrow 2/3 &
\xi_{i,n} \rightarrow \xi_{n} \\
X \rightarrow 1/2 & W \rightarrow 1/12 & \omega_{1} \rightarrow -2 &
\xi_{2} \rightarrow 2 & \xi_{1} \rightarrow -3/2 \\
& & & Y \rightarrow 1/2 & \omega_{2} \rightarrow 3,
\end{array}
\end{equation}
where the values for \(A\) and \(Z\) can be easily found using their
explicit form.

We can take a look at the partition function to see exactly
what quantities we have to know.  From (\ref{eq:partfun}), and using
(\ref{eq:scal}), we get
\begin{equation}
{{\cal Z}_{N+1} {\cal Z}_{N-1} \over {\cal Z}_{N}^{2}} =
(1+ {1 \over N}) R_{2N} R_{2N-1}
{W_{N} \over W_{N-1}}. \label{eq:relatezw}
\end{equation}
However, the ratio of \({\cal Z} \)'s is related to \(F''\),
so all we have to know is the ratio
\( W_{N} / W_{N-1} \), something that we can easily find with a
recursion relation for the \(W\)'s
(of course we also have to find the differential
equation satisfied by the function used in the \(W\) ansatz).

To find this recursion relation we only need eqs.(\ref{eq:equa24}),
(\ref{eq:equa26}), and (\ref{eq:equa27}).  After some algebraic
manipulations we get a recursion relation for \(W_{N} \),
\begin{eqnarray}
(- {(2N+1) \over \beta} + b R_{2N+1} ) ( - {2N \over \beta} + b
R_{2N}) [ 2 W_{N} - b R_{2N+1}] W_{N-1} \nonumber \\
+ R_{2N+1}(a+ b S_{2N})(a+ b S_{2N+1}) [ (- {2N \over \beta} W_{N-1})
+ {b^{2} \over 2} R_{2N} R_{2N-1}] = 0. \label{eq:finalw}
\end{eqnarray}
Using some information from table \ref{tablech2}, which applied for
recursion relations of \(P\)'s,
and using \( \phi = \mu = 2/5 \),
\( \rho = \nu = 1/5 \), we find the coefficients
listed in table~\ref{tablech4}.
\begin{table}[p]
\begin{center}
\begin{tabular}{|c|c|} \hline
order & eq.(\ref{eq:finalw}) \\ \hline
& \\
\( \beta^{-1/5} \) & \(0\) \\
& \\
\( \beta^{-2/5} \) & \( 6 f = h_{0}^{2} +  h_{0}' \) \\
& \\ \hline
\end{tabular}
\end{center}
\caption{Lowest order solutions of the recursion relation for the DIII --
even n -- matrix model.}
\label{tablech4}
\end{table}
Replacing solutions for \(R_{2N+l}\), and \(W_{2N+l}\) in (\ref{eq:relatezw}),
we get,
\begin{equation}
F'' = 4 f - h_{0}'.
\end{equation}
Using a power series expansion for \(f\) and replacing in
 \( 6 f=h_{0}^{2} + h_{0}' \), we find
\begin{equation}
h_{0} = \pm 2 \sqrt{3} t^{1/4} - {1 \over 8} t^{-1} \mp {5 \sqrt{3} \over
384} t^{-9/4} + \ldots.
\end{equation}
Finally, we end up with the following solution for the second derivative
of the free energy,
\begin{equation}
F'' = 4 f - h_{0}' \simeq 8 t^{1/2} \mp {\sqrt{3} \over 2} t^{-3/4}
- {13 \over 96} t^{-2} + \ldots.
\end{equation}
The appearance of a term \(O(t^{-3/4}) \) reveals that this model
describes unoriented surfaces for sure, as well as oriented ones.
  To compare with previous model, we consider the ratio of
coefficients,
\begin{equation}
{c_{0} c_{2} \over c_{1} ^{2}} = {8 \times - {13 \over 96} \over (\mp
{\sqrt{3} \over 2})^{2} } = - {13 \over 9}.
\end{equation}
This is the same ratio as what was found in the first solution of the
previous section so it describes exactly the same physics.

\section{Conclusion}

We now summarize our two main results, with their implications.  Firstly,
for each of the models, the free energy (up to overall scalings)
was the same as other models
previously studied and using completely different matrix ensembles,
as revealed by the ratio \({c_{0} c_{2} \over c_{1}^{2}} = -13/9\)).  This
means the following: if these models were actually describing
 gravity coupled
to some other system, then we expect that different regulators would
introduce a dependence of the free energy on a coupling parameter.
Hence using different matrix ensembles would yield different results.
But this is not the case so our analysis confirms that all of these
models describe pure gravity including both oriented and unoriented
surfaces, as it was first assumed for previous solutions.

Secondly, although both models yield exactly the same result, there is
an additional solution for \(n\)-odd (ratio \(83/9\))
.  This solution differs from the other one only by the coefficient of
the torus/Klein bottle term.  So all ratios that do not involve this term
are the same in both solutions.
A similar result was found in \cite{Myerp:90} for QRSD matrices, where
one solution did not include odd Euler character surfaces.  In the
present case, the physical interpretation of the extra solution
 remains unclear.

So far, the C I generator ensemble remains unstudied (as are most of
the circular ensembles, which integrate over the entire
symmetric space \cite{Rav:91}).
An interesting observation is that measures for D III - \(n\) even - and
C I are similar to A II and A I respectively after a change of variables
\(y=x^{2}\).  The D III and C I generator ensembles are more complicated
to analyse though, because their range of integration is from \(0\) to
\(\infty\) instead of \(-\infty\) to \(\infty\), which causes the appearance
of boundary terms in the recursion relations.  Given that D III with \(n\)
even and A II yield the same physics in the double--scaling limit, we
might expect C I and A I to do as well.

\section*{Acknowledgment}
I would like to thank Robert C. Myers, for his help throughout the
completion of this work.  This paper is a shortened version of a thesis
presented
in partial fullfilment for the degree of master in physics.  This research was
supported by NSERC.


\end{document}